\documentclass{article}

\usepackage{arxiv}

\usepackage[utf8]{inputenc} 
\usepackage[T1]{fontenc}    
\usepackage{hyperref}       
\usepackage{url}            
\usepackage{booktabs}       
\usepackage{nicefrac}       
\usepackage{microtype}      
\usepackage{lipsum}

\usepackage{cite}
\usepackage{amsmath,amssymb,amsfonts}
\usepackage{algorithmic}
\usepackage{graphicx}
\usepackage{textcomp}
\usepackage{xcolor}

\title{Generalised Quantum Tree Search}

\author{
  Andr\'{e} Sequeira\\
  Department of Mathematics\\
  University of Aveiro\\
  Aveiro, Portugal \\
  \texttt{andresequeira401@gmail.com} \\
   \And
Luis Paulo Santos\\
International Iberian Nanotechnology Laboratory\\
CSIG, INESC TEC\\
University of Minho\\
Braga, Portugal \\
\texttt{psantos@di.uminho.pt}
\And
Luis Soares Barbosa\\
International Iberian Nanotechnology Laboratory\\
HASLab, INESC TEC\\
University of Minho\\
Braga, Portugal \\
\texttt{lsb@di.uminho.pt}
}

\begin{document}
\maketitle

\begin{abstract}
This extended abstract reports on on-going research on quantum algorithmic approaches to the problem of generalised tree search that may exhibit effective quantum speedup, even in the presence of non-constant branching factors. Two strategies are briefly summarised and current work outlined.
\end{abstract}

\keywords{Quantum Tree Search, Quantum Speedup}

\section{Problem statement}
Knowledge representation has long been a cornerstone in Artificial Intelligence. Goal-based agents, also known as problem-solving agents use atomic representations to further reason about what actions and states to consider, given a goal. When the agent deals with known and deterministic environments, the process of reaching the goal can be formalised as a search problem \cite{Russel2012}. The possible action sequences starting at the initial state form a search tree where branches are actions and the nodes correspond to states in the state space of the problem. Therefore, reasoning reduces to tree searching. Quantum search algorithms, over some variant of the original Grover's Algorithm \cite{Grover1996}, perform search in superposition and have the potential to provide a quadratic speed-up. When the problem that the agent is trying to solve has a constant branching factor, Grover's algorithm can perform better than any classical \textit{uninformed} tree search algorithm \cite{Tarrataca2011}, even in an iterative deepening strategy \cite{Tarrataca2013}. However, all possible quantum speed-ups work under the weak assumption of a constant branching factor problem-solving agent. When we relax the assumption for non-constant branching factor problems, then, as proved in \cite{Tarrataca2011}, quantum algorithms no longer guarantee superior performance compared to classical tree search algorithms. This is due to the fact that in these cases, for classical algorithms the \textit{effective} branching factor converges to the \textit{average} one. However, in the quantum setting one must always use the \textit{maximum} branching factor to encode superpositions. This limiting factor results in certain cases in a slowdown compared to the classical counterpart. Real-world world problems, for example, route-finding problems \cite{Srinivasan2018}, typically have large, and most importantly, non-constant branching factors.\\
\textit{Uninformed} strategies are only of interest when dealing with small problems. In Artificial Intelligence, the search space typically shows exponential growth, so, it becomes intractable to perform an exhaustive search as in \textit{uninformed} tree search algorithms. In most of the real-world problems which we don't know the solution, it is still possible, by reasoning over the natural problem formulation, to assign values to particular states. Empowering \textit{informed} strategies i.e designing \textit{heuristic} functions, enable the search to be focused on promising nodes rather than uniformly expanding the search tree. This way, an agent reaches a solution typically in less time and consuming less memory. Classical \textit{informed} strategies can overcome the quadratic speed-up employed by Grover's algorithm. It is of great importance the study of quantum heuristics and their potential to scale quantum tree search algorithms.
This abstract reports on-going work on two novel strategies for performing tree search in the quantum setting: \textbf{(1)} generalisation of quantum tree search for both constant and non-constant transition models and \textbf{(2)} development of a novel \textit{informed} quantum tree search algorithm.
\section{Research directions}
\subsection{Generalising for non-constant branching}
In \cite{Tarrataca2011} the authors proposed a quantum 
algorithm for performing tree search. However, for non-constant branching factors, the quantum algorithm still needs to use the maximum branching factor, often leading to a slowdown compared to the classical counterpart. We aim to generalise the algorithm to deal with arbitrary non-constant branching factors. Rather than considering only the Hilbert space represented by the superposition of the actions at each level of the tree, consider having separate basis states representing the actions, a, and the state, s, in which the agent is, i.e. a node in the tree. Suppose that the search tree has action space $A$. For a depth $d$, even with a non-constant branching factor $b$, there will be at most $|A|^d$ leaf nodes. So, a node can be encoded in a quantum state with $log_2{|A|^d}$ qubits. Also represent an action initially in the ground state and a node that will be initially the root of the tree as the tensor product of both basis states: 

\begin{align}
            |s\rangle = |0\rangle^{\otimes{log_2{|A|^d}}}& , \quad |a\rangle = |0\rangle^{\otimes log_2{|A|}}\\
            |\psi_{0}\rangle &= |s\rangle \otimes |a\rangle
\end{align}

\begin{figure}[h]
            \centering
            \includegraphics[width=0.7\textwidth]{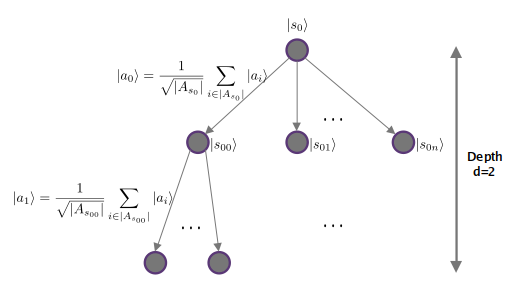}
            \caption{Superposition tree with arbitrary branching factor}
            \label{fig: superposition_tree}
        \end{figure}

        An unitary operator $\mathcal{A}$ may now be specified that prepares the superposition of the actions controlled by the state of the node. The superposition of admissible actions at a given node,$A_s$  
        \begin{equation}
            \mathcal{A} : |s\rangle \otimes |0\rangle^{\otimes{log_2{|A|}}} \mapsto |s\rangle \otimes {1 \over \sqrt{|A_s|}} \sum_{i \in |A_s|} |a_i\rangle
        \end{equation}
        The unitary operator $\mathcal{A}$ can be realised because zero amplitude can be maintained on the superposition terms that don't represent an admissible action \cite{ventura2000}. Another unitary operator $\mathcal{T}$ needs to be specified to handle state transitions, i.e. the evolution operator dealing with the traversal of the tree. 
        \begin{equation}
            \mathcal{T}: |s\rangle \otimes |a\rangle \mapsto |s'\rangle \otimes |a\rangle
        \end{equation} 
        These two unitary operators will be interleaved for a predefined depth $d$, at each iteration extending the initial Hilbert space with new basis states for representing a new action corresponding to the level of the tree. To reduce the complexity associated with the application of these operators,  only neighbouring states might be considered. Applying the operators $d$ times can be interpreted as computing the tree in superposition for a depth $d$, \autoref{fig: superposition_tree}.
        At this point a Grover oracle $O$, needs to be constructed, that reckons an arbitrary state as a goal state, inverting the associated phase:
        \[ 
                O|\psi\rangle = \left\{
                \begin{array}{ll}
                    - |s\rangle|a_{0}a_{1}...a_{d}\rangle\ \quad if \quad |s\rangle \mapsto goal state\\
                    |s\rangle|a_{0}a_{1}...a_{d}\rangle\ \quad otherwise\\
                \end{array} 
                \right. 
        \]
        Now, the Grover iterate will amplify the correct sequence of actions that lead to the goal state as intended. Interestingly enough, the Grover oracle in this case has a peculiar meaning. Indeed, the oracle inverts the phase of the sequence of actions that lead to the goal state. This means that the oracle reckons only the final state as the goal state, which is known. Previous works use oracles that reckons the correct sequence of actions. Therefore the oracle is viewed as an entity that already knows the correct sequence of actions to a given problem. This formulation ended up consuming more memory, due to the binary representation of the node. However, providing that the unitary operators $\mathcal{A}$ and $\mathcal{T}$ have an efficient representation, i.e. can be constructed in polynomial time, then the complexity of the algorithm will be dominated by the dimension of the search space. Thus Grover's algorithm will guarantee a quadratic speedup. This approach provides a way of dealing with arbitrary search trees with non-constant branching factors because Grover's algorithm will always have the correct search space associated. This algorithm can also work as an iterative deepening version of \cite{Tarrataca2013}.
        \subsection{Informed Quantum Tree Search} \label{subsec: informed-quantum}
        Traditionally, heuristics are employed to choose among possible tree paths, ideally producing an optimal sequence of actions resulting from a non-uniform expanded search tree. At first sight such an operation seems impossible to perform quantum mechanically. However, combining Amplitude Amplification \cite{Brassard2002} with some performance metric may yield a quantum heuristic that can perform some interesting tricks. It is possible to amplify terms in a general superposition state that respects some threshold value, or even perform \textit{pruning}. The latter case can be pictured by the rather trivial example of a 2-qubit uniform superposition state. In this case, if one marks a single superposition term then it is known that a single amplitude amplification iteration results in the marked state with certainty \cite{Boyer1998}. If instead of measuring the state, one keeps evolving it according to some operator, then it can be interpreted as pruning branches of a tree. 
        
\section{Current work}
The current work is being developed in both research directions discussed above. For \textbf{(1)} we seek a formal proof of the quadratic enhancement for non-constant trees. The algorithm is also being tested in various goal-based agents, and simulated in IBM’s quantum platform \cite{Qiskit2019}. For \textbf{(2)} the quantum pruning strategy is being applied in the context of informed tree search for an arbitrary number of qubits scheme. Moreover, a quantum greedy best-first algorithm that uses amplitude amplification and exponential search as subroutines is being developed.
        
\section*{Aknowledgement}
This research is financed by the ERDF through the Operational Programme for Competitiveness and Internationalisation - COMPETE 2020 Programme and by National Funds through the Portuguese funding agency, FCT, within project POCI-01-0145-FEDER-030947.

\bibliographystyle{unsrt}  

\end{document}